\documentclass[conference]{IEEEtran}
\renewcommand{\baselinestretch}{0.975}
\IEEEoverridecommandlockouts
\usepackage{cite}
\usepackage{amsmath,amssymb,amsfonts}
\usepackage{algorithmic}
\usepackage{algorithm}
\usepackage{graphicx}
\usepackage{textcomp}
\usepackage{xcolor}
\usepackage{booktabs}
\usepackage{academicons}
\usepackage{url}
\definecolor{orcidlogocol}{HTML}{A6CE39}
\usepackage{multirow}
\usepackage{tabularx}
\usepackage{makecell}
\usepackage{tabularx}
\usepackage{subcaption}
\usepackage{comment}
\usepackage{array}
\usepackage{threeparttable}

\begin{document}

\title{ECG Unveiled: Analysis of Client Re-identification Risks in Real-World ECG Datasets}

\author{
    \IEEEauthorblockN{
        Ziyu Wang\IEEEauthorrefmark{1},
        Anil Kanduri\IEEEauthorrefmark{2},
        Seyed Amir Hossein Aqajari\IEEEauthorrefmark{1},
        Salar Jafarlou\IEEEauthorrefmark{1},
        \\ Sanaz R. Mousavi\IEEEauthorrefmark{3},
        Pasi Liljeberg\IEEEauthorrefmark{2},
        Shaista Malik\IEEEauthorrefmark{1},
        and Amir M. Rahmani\IEEEauthorrefmark{1}
    }

     \IEEEauthorblockA{
        \IEEEauthorrefmark{1}University of California, Irvine, USA, \IEEEauthorrefmark{2}University of Turku, Finland, \IEEEauthorrefmark{3}California State University, Dominguez Hills, USA\\
        \{ziyuw31, saqajari, jafarlos, smalik, amirr1\}@uci.edu,  \{spakan, pasi.liljeberg\}@utu.fi, srahimimoosavi@csudh.edu
    }
}

\maketitle



\begin{abstract}
While ECG data is crucial for diagnosing and monitoring heart conditions, it also contains unique biometric information that poses significant privacy risks. Existing ECG re-identification studies rely on exhaustive analysis of numerous deep learning features, confining to ad-hoc explainability towards clinicians decision making. In this work, we delve into explainability of ECG re-identification risks using transparent machine learning models. We use SHapley Additive exPlanations (SHAP) analysis to identify and explain the key features contributing to re-identification risks. We conduct an empirical analysis of identity re-identification risks using ECG data from five diverse real-world datasets, encompassing 223 participants. By employing transparent machine learning models, we reveal the diversity among different ECG features in contributing towards re-identification of individuals with an accuracy of 0.76 for gender, 0.67 for age group, and 0.82 for participant ID re-identification. Our approach provides valuable insights for clinical experts and guides the development of effective privacy-preserving mechanisms. Further, our findings emphasize the necessity for robust privacy measures in real-world health applications and offer detailed, actionable insights for enhancing data anonymization techniques.
\end{abstract}


\section{Introduction}

The digitization of health records and the proliferation of wearable biosensors have revolutionized e-health~\cite{alikhani2024seal}, enabling continuous monitoring and real-time analysis of physiological signals~\cite{alikhani2024seal, aqajari2024enhancing}. Among these, Electrocardiogram (ECG) signals capture the heart's electrical activity through distinct PQRST complexes, providing vital insights into heart health and enabling the detection of various cardiac abnormalities~\cite{berkaya2018survey}. While ECG signals are primarily used for medical diagnosis and treatment, they have unique biometric properties that can be exploited to identify individuals~\cite{ghazarian2021increased}. 

\begin{figure}[tb]
    \centering
    \includegraphics[width=0.7\columnwidth]{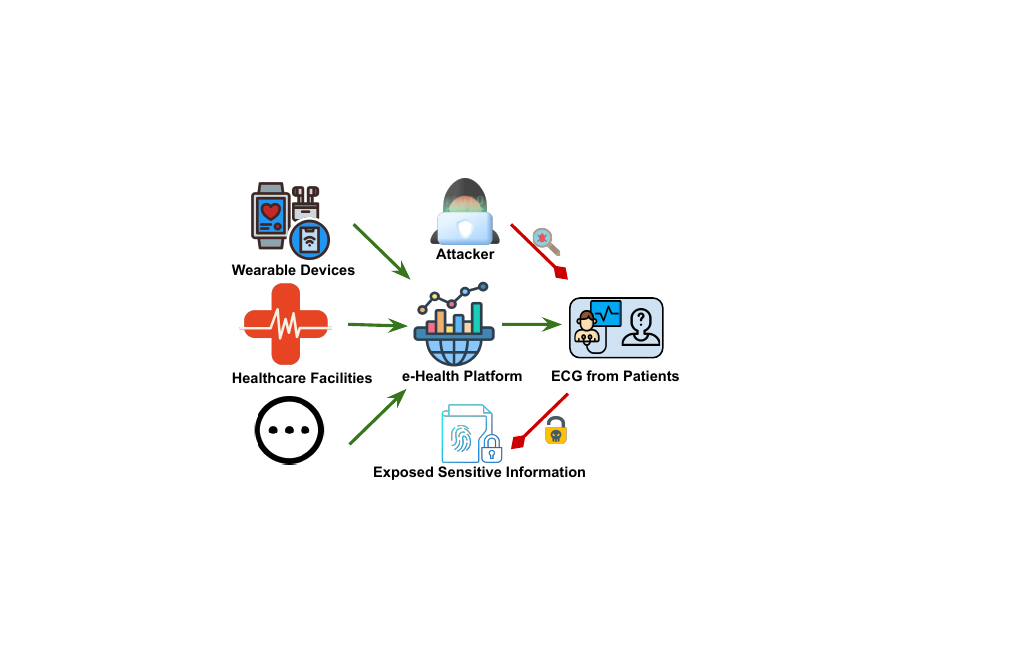}
    \vspace{-4pt} 
    \caption{Overview of the threat model illustrating ECG data aggregation from various sources to e-health platforms, creating an attack surface for potential client re-identification.}
    \vspace{-5pt}
    \label{fig:overview}
\end{figure}

As ECG data becomes more accessible through e-health platforms and health record databases, the risk of client re-identification from public datasets significantly increases~\cite{odinaka2012ecg, Wang_EMBC2024}. Public datasets are essential for research in healthcare. These datasets can be exploited through various machine learning-based attacks, such as linkage attacks~\cite{powar2023sok} and membership inference attacks~\cite{shokri2017membership}, further exacerbating client re-identification risks (Fig. \ref{fig:overview}). Attackers can leverage the distinct biometric features of ECG signals to re-identify clients, leading to significant privacy breaches~\cite{yang2022zebra} and potential misuse of personal health information. This highlights the urgent need for robust privacy-preserving mechanisms to safeguard clients' identities in publicly accessible datasets~\cite{plataniotis2006ecg}.

Recent studies have shown that ECG signals can be used for biometric authentication, revealing privacy risks as ECG statistical variants can disclose individual identities~\cite{ghazarian2021increased}. These vulnerabilities allow for pinpointing individuals within subgroups based on demographic information and health conditions~\cite{ingale2020ecg, yadav2018evaluation}. Similar privacy risks exist in other biosignal data types, such as Photoplethysmograph (PPG) and Electroencephalogram (EEG), highlighting the need for robust privacy-preserving techniques across all biosignal health systems~\cite{wang2020guardhealth, yao2020privacy}. The unique variants of ECG signals, illustrated by the PQRST peaks and key features (Fig. \ref{fig:pqrst_features}), can be exploited for client re-identification. Existing studies often rely on homogeneous datasets and controlled conditions that fail to capture the diversity and complexity of practical applications~\cite{ingale2020ecg}. For example, homogeneous datasets may consist of ECG recordings from a single demographic group or clinical setting, while controlled conditions involve standardized environments that do not reflect everyday variability. In contrast, real-world data is inherently diverse, covering various demographic groups, clinical conditions, and recording environments. Therefore, a comprehensive study and analysis of client re-identification risks in diverse, real-world ECG datasets is needed to better understand and mitigate these risks.

\subsection*{\textbf{Research Gaps and Our Contributions}}

The primary challenges in current research include the following: (i) existing research has not adequately considered real-world threat models, limiting their applicability~\cite{ghazarian2021increased}; (ii) the reliance on deep learning models is problematic because these models extract latent space features that lack explainability and cannot be interpreted by humans~\cite{cheng2024efflex}, leaving domain experts without the necessary information to develop effective privacy-preserving mechanisms~\cite{ras2022explainable}; and (iii) previous works did not thoroughly investigate the factors contributing to client re-identification risks. Their methods often rely on trial experiments or observations on small samples, leading to inadequate and inefficient examination of key features~\cite{ingale2020ecg}.

In this study, we address these gaps by investigating the re-identification risks associated with ECG data using transparent machine learning models. A key aspect of our approach is the use of SHAP analysis to identify the most influential features contributing to re-identification. By building interpretable models, we evaluate the effectiveness of ECG signals in re-identifying individuals across diverse demographic groups and clinical conditions within real-world heterogeneous datasets. This analysis provides valuable insights that can guide the development of privacy-preserving techniques and inform clinical experts.

In summary, our contributions are as follows:
\begin{itemize}
    \item We provide a comprehensive analysis of re-identification risks using transparent machine learning models, applied to heterogeneous datasets from multiple sources, to accurately reflect real-world scenarios.
    \item We offer findings that highlight the need for robust anonymization techniques and privacy-preserving mechanisms, providing a foundation for future research aimed at protecting sensitive health information.
    \item We conduct a rigorous investigation into the factors contributing to re-identification, utilizing feature importance assessment and SHAP analysis to provide insights that are interpretable by domain experts.
\end{itemize}


\section{Analysis of Re-identification Risks}

\subsection{Method}

Our primary objective is to assess the re-identification risks associated with ECG data. To achieve this, we extract key PQRST features, which represent distinct electrical activities of the heart and are crucial for identifying individual-specific patterns~\cite{hurst1998naming}. Focusing on these features allows us to build interpretable models that reveal re-identification factors and provide actionable insights for clinical experts.

\subsubsection{Feature Extraction from ECG Signals}


\begin{figure}[ht]
    \centering
    \vspace{-6pt} 
    \includegraphics[width=0.78\linewidth]{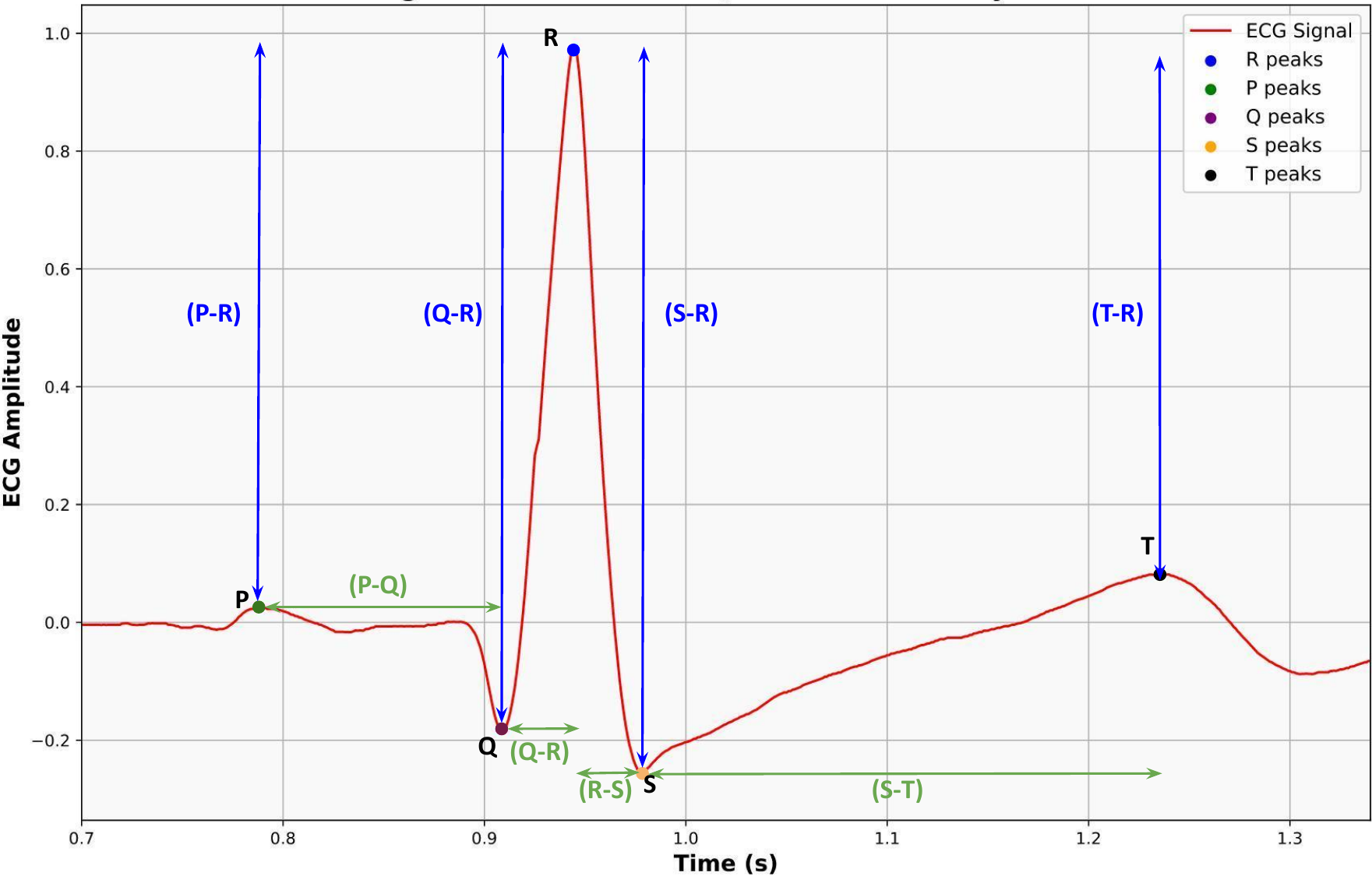}
    \vspace{-5pt} 
    \caption{ECG signal with detected PQRST peaks and key features, highlighting its biometric properties.}
    \label{fig:pqrst_features}
    \vspace{-5pt} 
\end{figure}

We extracted key PQRST features from ECG signals using the NeuroKit2 ~\cite{Makowski2021neurokit} library. The process involved three main steps: ECG signal cleaning, R-peak detection, and delineation of P, Q, S, and T peaks to identify precise fiducial points. Signal cleaning was essential to address noise and baseline wander in raw ECG signals. We applied a highpass Butterworth filter to remove slow drifts and direct current (DC) bias, followed by a powerline filter to eliminate 50 Hz interference from electrical sources.


To capture distinctive ECG characteristics for identifying patterns associated with individuals, we derived several statistical features, such as mean amplitude differences and interval durations. Previous studies have shown that these statistical features effectively distinguish individual-specific cardiac patterns in the PQRST complex, making them suitable for re-identification tasks~\cite{ghazarian2021increased, ingale2020ecg}. Specifically, the P, Q, R, S, and T peaks represent different phases of the cardiac cycle and are critical for capturing the heart's electrical activity~\cite{hurst1998naming}. These peaks can be used to extract clinically significant features.

We computed the Mean Amplitude Differences between the P, Q, S, T peaks and the R-peak:
\vspace{-8pt}
\begin{equation}
\vspace{-2pt}
\text{Mean Amplitude Difference (X-R)} = \frac{1}{N} \sum_{i=1}^{N} (A_{X,i} - A_{R,i})
\end{equation}
where \(X\) represents the P, Q, S, or T peaks, \(A_X\) and \(A_R\) are the amplitudes at these peaks, and \(N\) is the number of valid peak pairs. Additionally, we calculated the Mean Intervals between the clinically significant P, Q, R, S, and T peaks to capture temporal features. These include:
\vspace{-10pt}

\begin{equation}
\text{Mean Interval (X-Y)} = \frac{1}{N} \sum_{i=1}^{N} (t_{Y,i} - t_{X,i})
\end{equation}

where \(X\) and \(Y\) represent different peaks (i.e., P, Q, R, S, T), and \(t_X\) and \(t_Y\) are the times of these peaks.


As illustrated in Fig. \ref{fig:pqrst_features}, we identified and highlighted the key PQRST features on an ECG signal. The figure shows the amplitude differences (e.g., \(P-R\), \(Q-R\), \(S-R\), \(T-R\)) and intervals (e.g., \(P-Q\), \(Q-R\), \(R-S\), \(S-T\)) between significant points. Unlike deep learning models, which often generate features in latent spaces that are difficult to interpret and understand \cite{lipton2018mythos}, these explainable features capture unique physiological variations in heart activity among individuals, considering demographic, health, lifestyle, and genetic factors, which are particularly effective for re-identifying individuals~\cite{vecht2009ecg}.

\begin{table*}[!ht]
\vspace{-6pt}
\centering
\caption{Summary of ECG Datasets Used in Experiments}
\vspace{-5pt}
\label{tab:ecg_datasets}
\begin{tabular}{|l|c|c|c|c|c|}
\hline
\textbf{Dataset} & \textbf{Subjects} & \textbf{Age Range} & \textbf{Gender (M/F)} & \textbf{Sampling Rate (Hz)} & \textbf{Health Condition} \\ \hline
MIT-BIH Arrhythmia~\cite{moody2001impact} & 47 & 23-89 & 25/22 & 360 & Arrhythmias \\ \hline
SHAREE~\cite{melillo2015automatic} & 139 & 55-72 & 90/49 & 128 & Hypertension \\ \hline
BIDMC CHF~\cite{baim1986survival} & 15 & 22-71 & 11/4 & 250 & Congestive Heart Failure \\ \hline
Brno University of Tech~\cite{nemcova2020brno} & 15 & 21-83 & 9/6 & 1000 & General population \\ \hline
MIT-BIH Long-Term~\cite{moody2001impact} & 7 & 46-88 & 6/1 & 128 & Long-term general monitoring \\ \hline
\textbf{Combined} & \textbf{223} & \textbf{21-89} & \textbf{141/82} & \textbf{Multiple} & \textbf{Multiple} \\ \hline
\end{tabular}

\vspace{-8pt}
\end{table*}
\subsubsection{Data Processing}

Our experiments targeted three main tasks: binary gender re-identification, age group re-identification, and participant ID re-identification. These tasks were chosen because they are common targets in privacy leakage practices, making understanding their re-identification risks crucial~\cite{haas2011aspects}. Binary gender and age group re-identification are essential due to their frequent use in demographic analyses and targeted services, where age-related information is often sensitive \cite{yao2020privacy, haas2011aspects}. Participant ID re-identification assesses the risk of uniquely identifying individuals within a dataset, highlighting the privacy implications in longitudinal data \cite{odinaka2012ecg}. The data splitting methodologies for these tasks are summarized in Tab. \ref{tab:data_splitting}.

\begin{table}[!ht]
\centering
\caption{Data Splitting Methods for Re-identification Tasks}
\vspace{-5pt}
\label{tab:data_splitting}
\begin{threeparttable}
\begin{tabular}{|l|c|c|l|}
\hline
\textbf{Target} & \textbf{Train} & \textbf{Test} & \textbf{Label} \\ \hline
Gender & 80\% participants & 20\% participants & Binary \\ \hline
Age Group & 80\% participants & 20\% participants & Age group\tnote{1} \\ \hline
Participant & First 80\% of each & Last 20\% of each & Unique \\ 
ID & participant's data & participant's data & ID \\ \hline
\end{tabular}
\begin{tablenotes}
\item[1] Age groups: 21-30, 31-40, 41-50, 51-60, 61-70, 71-89 years.
\vspace{-10pt}
\end{tablenotes}
\end{threeparttable}
\end{table}

\subsection{Experiment} \label{sec.results}



\noindent \textbf{Model Training}
To investigate re-identification risks in ECG data, we used interpretable and explainable models including \textit{logistic regression}, \textit{decision trees}, and \textit{random forest}. Logistic regression offers straightforward coefficient interpretation, decision trees provide clear decision paths, and random forests offer feature importance scores~\cite{linardatos2020explainable}. SHAP works well with these models by attributing each feature's contribution to the prediction, providing consistent and locally accurate explanations~\cite{lundberg2017unified}. This approach helps identify key features and provides actionable insights for clinical experts, ensuring the safe use of ECG data while maintaining privacy.

\noindent \textbf{Model Evaluation and Interpretability Analysis} The tuned models were evaluated using standard classification metrics, including accuracy, precision, recall, F1-score, and ROC AUC, and confusion matrices were generated to visualize performance across different classes. We conducted SHAP analysis to understand the contributions of each feature to the model’s predictions. Let \(f\) be the model, \(x\) be the feature vector, and \(x_i\) be the value of the \(i\)-th feature. SHAP values \(\phi_i(x)\) represent the contribution of \(x_i\) to the prediction \(f(x)\):
\vspace{-5pt}
\begin{equation}
f(x) = \phi_0 + \sum_{i=1}^{M} \phi_i(x)
\vspace{-4pt}
\end{equation}
where \(\phi_0\) is the base value (mean prediction) and \(M\) is the number of features.


\noindent \textbf{Dataset}
In our experiments, we utilized five distinct real-world ECG datasets, each offering a diverse range of demographic and clinical conditions (refer to Tab. \ref{tab:ecg_datasets}). For each dataset, we extracted appropriate segments (30-120 minutes) of ECG recordings to capture continuous data and reflect real-world scenarios. This diverse, multi-sourced approach enhanced the robustness of our assessment of re-identification risks associated with ECG data.

\subsection{Results}

\subsubsection{Re-identification Risk}


\begin{table}[h]
\centering
\caption{Performance Metrics for Re-identification Models}
\vspace{-5pt}
\label{tab:reidentification}
\begin{tabular}{|l|c|c|c|}
\hline
\textbf{Re-identification Task} & \textbf{Accuracy} & \textbf{Precision} & \textbf{F1} \\ \hline
Gender & 0.755 & 0.766 & 0.760 \\ \hline
Age Group & 0.671 & 0.623 & 0.633 \\ \hline
Participant ID & 0.819 & 0.817 & 0.810 \\ \hline

\end{tabular}

\vspace{-10pt}
\end{table}

The models were trained and validated on data from the multi-sourced datasets, comprising 223 participants. The gender re-identification model achieved an accuracy of 0.755, the age group re-identification model achieved an accuracy of 0.671, and the participant ID re-identification model achieved an accuracy of 0.819. Detailed performance metrics are presented in Tab. \ref{tab:reidentification}.

High accuracy in gender and age group classification indicates that even without access to the target client's specific data during training, it is possible to infer the data owner's age range and gender using a small segment of ECG data or statistical features. This capability poses a significant threat to privacy, as adversaries can categorize individuals based on demographic attributes from minimal data. For participant ID re-identification, the model's high accuracy (0.819) suggests that an attacker can confidently match a small segment of a client's ECG data to their identity among 223 participants. This poses a severe threat to e-health systems, as depicted in Fig. \ref{fig:overview}, facilitating unauthorized identification and potential misuse of personal health information.

These privacy breaches can lead to unauthorized access to sensitive health data, discrimination based on health status, and loss of trust in e-health systems. E-health systems have increasingly relied on machine learning enhancements in recent years to improve diagnostics, personalize treatment plans, and predict patient outcomes~\cite{Wang_EMBC2024, wang2020guardhealth}. Consequently, these systems often utilize Machine Learning as a Service (MLaaS) to manage large datasets and deploy complex models efficiently. However, unauthorized information obtained through re-identification can be combined with other attacks, such as attribute inference and membership inference attacks, further compromising privacy~\cite{shokri2017membership}. As highlighted by studies on linkage attacks and profiling attacks~\cite{powar2023sok}, these risks threaten the integrity of MLaaS systems. Therefore, robust privacy-preserving techniques are crucial to mitigate these risks and protect individual privacy when deploying ECG data in clinical and research settings.

\subsubsection{Re-identification Factors}

\begin{figure*}[!ht]
    \centering
    \begin{subfigure}[b]{0.325\textwidth}
        \centering
        \includegraphics[width=\textwidth]{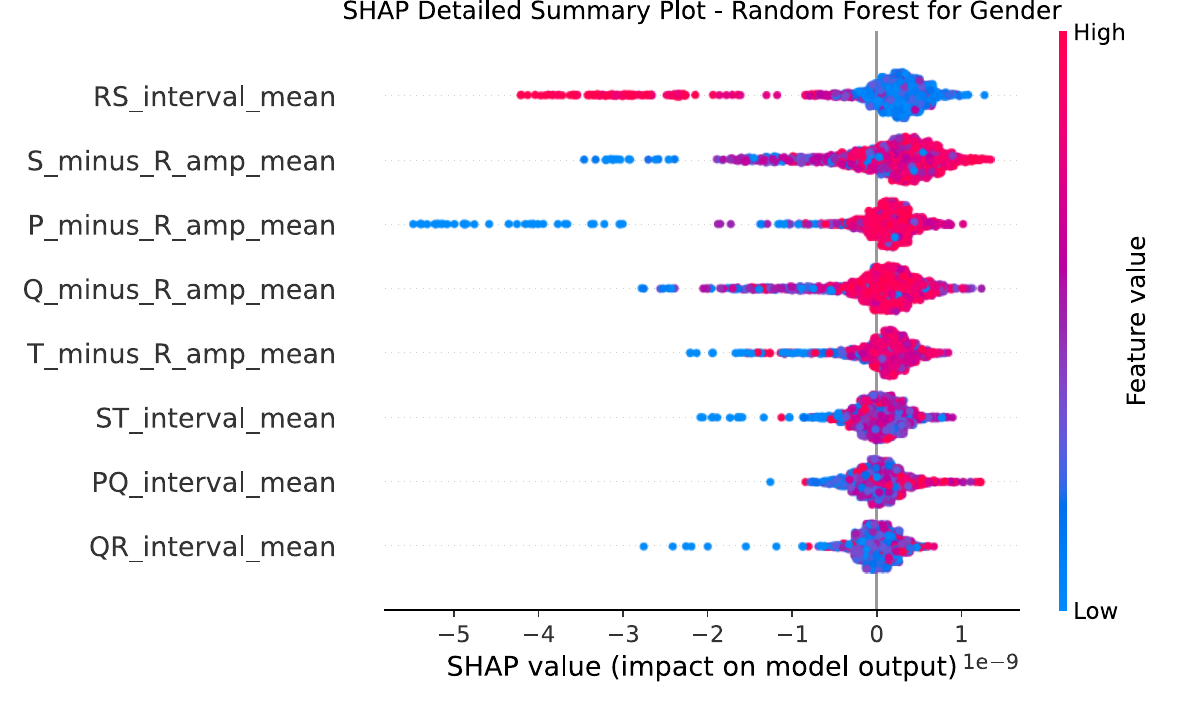}
        \caption{Gender Re-identification}
        \label{fig:gender_reidentification}
    \end{subfigure}
    \hfill
    \begin{subfigure}[b]{0.325\textwidth}
        \centering
        \includegraphics[width=\textwidth]{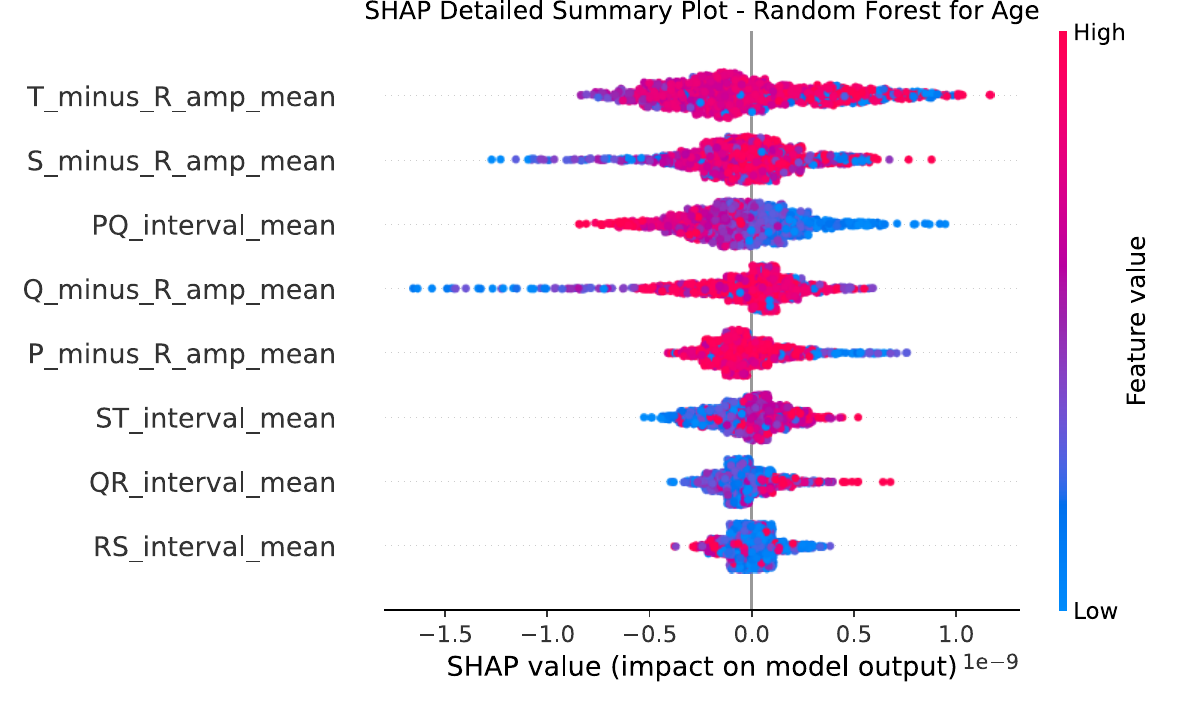}
        \caption{Age Group Re-identification}
        \label{fig:age_reidentification}
    \end{subfigure}
    \hfill
    \begin{subfigure}[b]{0.325\textwidth}
        \centering
        \includegraphics[width=\textwidth]{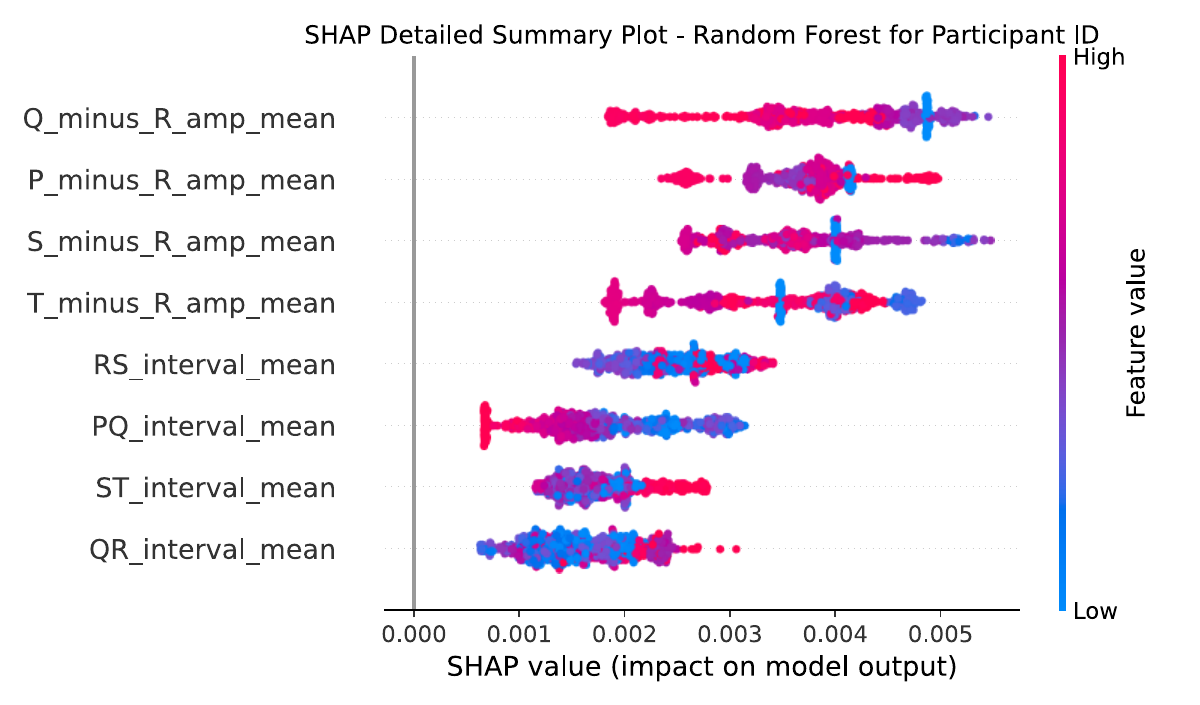}
        \caption{Participant ID Re-identification}
        \label{fig:participant_id_reidentification}
    \end{subfigure}
    \caption{SHAP Analysis for Re-identification Tasks.}
    \vspace{-5pt}
    \label{fig:shap_analysis}
    \vspace{-12pt}
\end{figure*}

The combined analysis, shown in Fig. \ref{fig:shap_analysis}, reveals the factors contributing to re-identification risks and offers actionable insights for clinical experts. Notably, features such as S-R and P-R amplitude differences were consistently significant across tasks, indicating their strong influence on re-identification.

For gender re-identification (Fig. \ref{fig:shap_analysis}a), the R-S interval and S-R amplitude difference were prominent. Clinically, the R-S interval, representing the time between the peak of the R wave and the end of the S wave, varies significantly due to gender differences in cardiac structure and function~\cite{hurst1998naming}. Larger S-R amplitude difference can indicate variations in ventricular depolarization, which often differ between men and women due to anatomical and physiological differences in the heart~\cite{ vecht2009ecg}. Additionally, differences in P-R amplitude difference reflect anatomical differences, which can be influenced by gender-specific factors such as hormonal effects on the autonomic nervous system.

In age group re-identification (Fig. \ref{fig:shap_analysis}b), the T-R amplitude and P-Q interval played crucial roles. Age-related changes in the cardiovascular system, such as increased arterial stiffness and altered atrial conduction, affect these intervals~\cite{vecht2009ecg}. The variability in T-R amplitudes among different age groups reflects these physiological changes. The P-Q interval, which measures atrial to ventricular conduction time, also varies with age due to structural and functional changes in the heart's conduction system. For participant ID re-identification (Fig. \ref{fig:shap_analysis}c), Q-R amplitude difference and P-R amplitude difference were particularly impactful. The Q-R amplitude represents the voltage difference between the Q and R waves, which can vary significantly among individuals due to differences in myocardial mass and conduction pathways. Also, the P-R amplitude difference highlights individual variations in atrial depolarization and ventricular depolarization dynamics~\cite{vecht2009ecg}. These insights help address privacy concerns by identifying critical ECG attributes that need protection. Understanding how specific ECG features contribute to re-identification allows for the development of more secure and transparent biometric systems, safeguarding individual privacy while utilizing ECG data for clinical and research purposes.
\vspace{-3pt}
\section{Conclusion}
\vspace{-3pt}
This study comprehensively analyzed the re-identification risks associated with ECG data using traditional statistical features and transparent machine learning models. By validating our approach across five diverse datasets, we demonstrated that ECG signals contain sufficient biometric information to significantly compromise privacy, achieving high accuracy in re-identifying individuals. Through SHAP analysis, we identified the most impactful features, providing critical insights for clinical experts and guiding the development of effective anonymization techniques. These findings highlight the urgent need for robust privacy-preserving mechanisms to safeguard patient biosignal data in real-world health applications.


\bibliographystyle{unsrt}
\bibliography{ref_compact}
\end{document}